\newcommand{\ee}{e^{+}e^{-}}

\newcommand{\jp}{J/\psi}

\newcommand{\mumu}{\mu^{+}\mu^{-}}
\newcommand{\pipi}{\pi^{+}\pi^{-}}

\newcommand{\pim}{\pi^{-}}
\newcommand{\pip}{\pi^{+}}
\newcommand{\piz}{\pi^{0}}

\newcommand{\rt}{\rightarrow}

\newcommand{\etal}{\em et al.}

\documentclass{PoS}

\title{The Role of Flavor Physics in the LHC Era}

\ShortTitle{Flavor Physics in the LHC Era}

\author{\speaker{Stephen Lars Olsen}\thanks{Supported by the Korean National Research Foundation (NRF)
Grant No.~20110029457.    }
\footnote{Current address: Institute for Basic Science, Daejeon, Korea}
\\
        Seoul National University\\
        E-mail: \email{solsen@hep1.snu.ac.kr}}

\abstract{Although searches for new physics at the CERN Large Hadron Collider will probably
dominate the the agenda of the experimental high energy physics community during the next decade or more,
high-intensity experiments at the $\tau$-charm and beauty thresholds will continue to play important
complementary roles.  These include the establishment of stringent constraints on proposed theories
for beyond-the-Standard-Model physics and unique opportunities to address some new physics scenarios
that are inaccessible at the LHC.  In addition, in the event that the LHC does discover some new
phenomena, high sensitivity flavor physics measurements could provide diagnostic clues as to the   
physics processes responsible for the observed effects.  In this talk I present a few examples that
illustrate the close inter-relation of new physics searches at the high-energy frontier and 
high-sensitivity measurements at the intensity frontier.}

\FullConference{52 International Winter Meeting on Nuclear Physics - Bormio 2014,\\
		27-31 January 2014\\
		Bormio, Italy }

\begin{document}

\section{Introduction}

\noindent
The discovery of the Higgs boson in 2012 by the ATLAS~\cite{higgs_atlas} and CMS~\cite{higgs_cms} experiments
at the CERN Large Hadron Collider (LHC) provided the long-awaited experiment evidence that ``completed'' the
Sandard Model (SM).  Although this model has been phenomenally successful at reproducing the results of all
particle physics experiments that have been reported to date, it has a large number of parameters that have
to be input from experiment, it does not provide any explanation for dark matter, dark energy or the
matter-antimatter asymmetry of the Universe, and does not incorporate Gravity.
For these and other reasons, most particle physics researchers believe that the SM, in
spite of its considerable success, is not a complete theory of nature.

As a result, a number of extensions of the SM, so-called beyond-the-SM (BSM) theories, have been proposed,
most of which predict the existence of new, as yet unseen, massive particles~\cite{snowmass2013}, usually
with masses in the 100~GeV$\sim$2~TeV range that is accessible at the LHC.  The most commonly
discussed BSM theory is Supersymmetry (SUSY), which proposes a SUSY partner for each of the established
SM partners, and an assortment of five Higgs particles that include a doublet of charged Higgs scalars~\cite{susy}.
Other BSM theories predict heavy versions of the $Z$ and $W$ bosons~\cite{w-prime}, a fourth-generation of
quarks~\cite{t-prime}, etc.  To the disappointment of many practitioners in the field, the first operational
period of the LHC, which studied $pp$ collisions in the $\sqrt{s} = 7\sim 8$~TeV range, uncovered no new particles
(besides the Higgs), and most of the SUSY parameter space for SUSY partner particle
with masses below $M\sim 1$~TeV has been ruled out~\cite{LHC1_SUSY}.  However, the LHC is now in the process
of (approximately) doubling both the CM energy and the luminosity, and hopes remain high that first
signs of BSM particles will emerge in the higher energy data that will start becoming available in
early 2015.  It can be expected that the LHC and the ATLAS and CMS energy frontier, high-$p_T$, high luminosity
experiments will remain the ``Flagship'' experimental high energy particle physics programs for at least the
next decade.

In contrast to ATLAS and CMS, the LHCb experiment runs at lower luminosity (a few fb$^{-1}$/yr) and exploits the
large cross section for $B$ meson production in multi-TeV $pp$ collisions ($\simeq 300\mu$b at $\sqrt{s}=7$~TeV)
to do high precision measurements of the decay properties of particles containing $b$- and/or $c$-quarks.
The LHCb experiment has been remarkably productive in a number of areas.  It has recently made a
measurement (with $\sim4\sigma$ statistical significance) of the very rare decay $B_s\rt\mumu$ process
(${\mathcal B}(B_s\rt \mumu \simeq 3\times 10^{-9}$!)~\cite{LHCb_bs2mumu,uwer}\footnote{Similar
results from CMS were reported at this meeting by L.~Sonnenschein~\cite{CMS_bs2mumu,sonnenschein}.}
and made spectacular measurements
of the $B_s$-$\bar{B}_s$ mixing frequency with an impressive 0.1\%-level precision:
$f=\Delta m_s=17.768\pm 0.0.23\pm 0.006$ps$^{-1}$~\cite{LHCb_bsbsbar,uwer}.  As discussed at this meeting
by U.~Uwer~\cite{uwer}, these and other measurements place strong constraints on a number of proposed
BSM theories~\cite{bsmix-mumu_theory}.    

Prior to the startup of LHCb, the BaBar and Belle B-factory experiments dominated the landscape of $B$ and $D$
meson decay physics.  The main goals of these experiments were tests of the SM mechanism for CP violation~\cite{KM}.
Here the highlight of both groups' research programs in the first half of the 2000-2009 decade were measurements
of the $CP$ violating phase $\phi_1$ (aka $\beta)$ in time-dependent $CPV$ asymmetries in $B^0$ meson decays into
$CP$ eigenstates such as $K^0\jp$.  Their measurements~\cite{belle-babar_CPV} confirmed SM expectations and 
led to Nobel prizes for Kobayashi and Maskawa in 2008.   Subsequent notable results from these experiments,
which placed important constraints on BSM theories, included measurements of time-dependent $CP$ violating phases in penguin
dominated $CP$ eigenstate decay modes such as $B^0\rt K^0\phi$ and $B^0\rt K^0\eta'$~\cite{belle-babar_penguin}
and measurements of $B$ mesons decays to final states containing $\tau$ leptons, {\it i.e.},  purely leptonic
$B^+\rt \tau^+\nu_{\tau}$~\cite{belle_tau,babar_tau} and semileptonic 
$B\rt D^{(*)}\tau^+\nu_{\tau}$~\cite{belle_dtau,babar_dtau} decays.

BaBar stopped taking data in 2008 by which time it had accumulated a data sample that corresponded to a total
integrated luminosity of 531~fb$^{-1}$; Belle stopped in 2010 after accumulating a  1040~fb$^{-1}$ 
data sample.  The KEKB collider and the Belle detector are currently being upgraded to SuperKEKB and BelleII.  The
SuperKEKB design luminosity is 40 times that of KEKB and will start providing yearly data samples of 10~ab$^{-1}$ by about
2019 (1~ab$^{-1}$=~1000~fb$^{^1}$).  The LHCb and BelleII programs will be largely complementary: the yearly samples of $B$
mesons registered by LHCb and BelleII will be similar.   Because of the larger boost factors, LHCb will have
much better vertexing than BelleII and they will accumulate $B$ and $B_s$ decays simultaneously. On
the other hand, thanks to the clean $\ee\rt \Upsilon(4S)\rt B\bar{B}$ environment, BelleII will be uniquely
capable of studying inclusive modes such as $b\rt s\gamma$ and $b\rt u\gamma$, and modes with missing energy,
such as $B^+\rt\tau^{-}\nu_{\tau}$. However, for $B_s$ physics, BelleII would have to run at the $\Upsilon(5S)$,
where the production cross section for $B_s$ and $B$ meson production is about a factor of five lower. 

In this report I will give a few examples of how intensity-frontier flavor physics measurements impact
BSM searches at the energy-frontier.  For more comprehensive discussions I refer the reader to
Refs.~\cite{belleii_physics} and~\cite{superb_physics}.  

\section{Particle-antiparticle mixing}

\begin{figure}[htb]
\centerline{  \includegraphics[height=0.35\textwidth,width=0.7\textwidth]{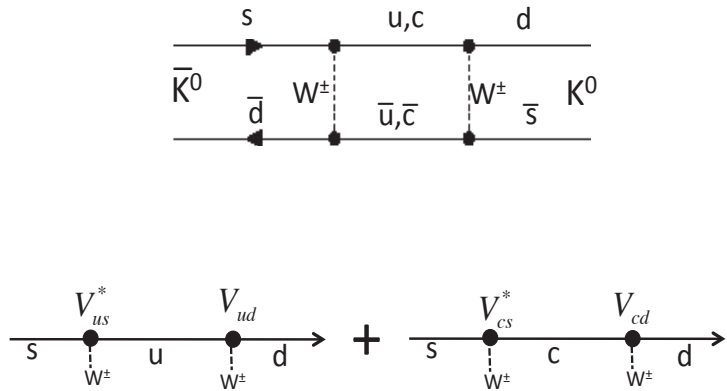}}
\caption{\footnotesize One of the quark-level box diagrams responsible for $K^0$-$\bar{K}^0$
mixing {\it (upper)}.  Illustration of the $\Delta m_K$ calculation
{\it (lower)}. 
}
\label{fig:k-mix}
\end{figure}

\noindent
The idea that neutral $K^0$ mesons would spontaneously change into their antiparticle $\bar{K}^0$ (and
{\it vice versa}) was first proposed by Gell Mann and Pais in 1955~\cite{pais-gellmann}.  Measurements
of the properties of the related $K_S$ and $K_L$ eigenstates was a major activity in the 1960s and led
to the discovery of $CP$ violation in 1964~\cite{fitch-cronin}.  In the SM, the frequency for $K^0$-$\bar{K}^0$
mixing or, equivalently, the $K_S$-$K_L$ mass difference $\Delta m_K$,  is adequately described by the imaginary
part of the four-quark process shown in Figure~\ref{fig:k-mix}.

From the lower portion of Fig.~\ref{fig:k-mix}, we can see that
\begin{equation}
\Delta m_K\propto G^2_F(V^{*}_{us}V_{ud}f(m_u)+V^*_{cs}V_{cd}f(m_c)).
\label{eq:Dms}
\end{equation}
In a four-quark world, the quark-flavor mixing matrix is simply a two-dimensional rotation
by the Cabibbo angle ($\theta_{C}$):
\begin{equation}
\left(
\begin{array}{cc}
V_{ud} & V_{us} \\
V_{cd} & V_{cs} 
\end{array} \right)
= \left(
\begin{array}{cc}
\cos\theta_{C}  & \sin\theta_{C} \\
-\sin\theta_{C} & \cos\theta_{C}
\end{array} \right),
\end{equation}
and Eq.~\ref{eq:Dms} becomes
\begin{equation}
\Delta m^{\rm SM}_K \propto G^2_F(f(m_u)-f(m_c))\cos\theta_{C}\sin\theta_{C}\simeq G^2_F m^2_c\cos\theta_{C}\sin\theta_{C}.
\end{equation}
The mixing frequency depends on the difference beiween the $c$- and $u$-quark masses, in fact, almost
entirely on the $c$-quark mass.  Thus,
when Glashow, Illiopulis and Maiani proposed the existence of the charmed quark in 1970~\cite{gim},
they predicted its mass to be ``not larger than 3$\sim$4 GeV,'' based on the measured value of $\Delta m_K$
at that time.

\begin{figure}[htb]
\centerline{  \includegraphics[height=0.35\textwidth,width=0.7\textwidth]{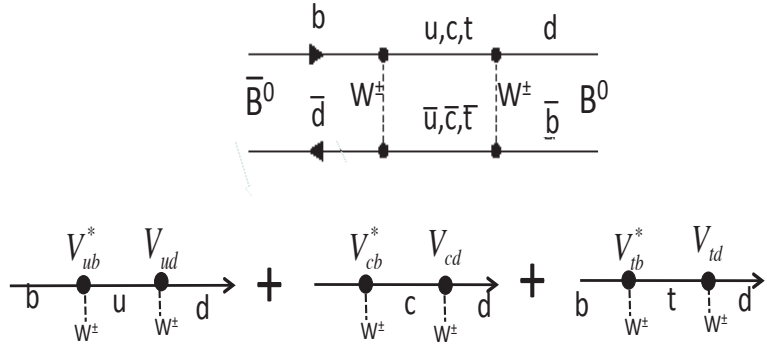}}
\caption{\footnotesize One of the quark-level box diagrams responsible for $B^0$-$\bar{B}^0$
mixing {\it (upper)} and an illustration of the $\Delta m_d$ calculation
{\it (lower)}. 
}
\label{fig:b-mix}
\end{figure}  

The major flavor physics event of of the 1980s was the unexpected discovery of $B^0$-$\bar{B}^0$ mixing by the
ARGUS experiment in 1987~\cite{argus}.  In this case, since the mixing involved the third-generation
$b$-quark, a six-quark analysis, as indicated in Fig.~\ref{fig:b-mix}, is necessary.  The
expression for the neutral $B$ meson eigenstate mass difference $\Delta m_d$ is
\begin{equation}
\Delta m^{\rm SM}_d\propto G^2_F(V^{*}_{ub}V_{ud}f(m_u)+V^*_{cb}V_{cd}f(m_c)+V^{*}_{tb}V_{td}f(m_t)),
\label{eq:Dmb}
\end{equation}
where $V_{ij}$ is  the $ij^{\rm th}$  element of the well known CKM six-quark-flavor mixing-matrix.
Note that if the quark masses were degenerate, {\it i.e.}, $m_u=m_c =m_t$, $\Delta m_d$
would be proportional to $V^{*}_{ub}V_{ud}+V^*_{cb}V_{cd}+V^{*}_{tb}V_{td}$, which the unitarity of
the CKM guarantees to be zero.  So, in this case also, the $B^0$-$\bar{B}^0$ mixing frequency
depends on the {\it non-degeneracy} of the quark masses and, to a good approximation depends
primarily on the top-quark mass. The SM expectation is
\begin{equation}
\Delta m^{\rm SM}_d\propto G^2_Fm^2_t|V^*_{tb}V_{td}|^2.
\label{eq:Dmb1}
\end{equation}
Thus, as a consequence of the ARGUS discovery of large $B`^0$-$\bar{B}^0$ mixing, it was realized that
the top-quark mass was much higher than was previously thought to be the case.\footnote{Much to the dismay
of those of us who, at the time, were searching for the $t$-quark in the $m_t \simeq 30$~GeV mass region.}
Specific calculations gave
values around $m_t\simeq 170$~GeV~\cite{mtop}.  The top quark was discovered in 1995~\cite{top-quark}
and its mass is measured to be 173.5~GeV with $\simeq$0.3\% precision~\cite{PDG}.

\subsection{Influence of SUSY on particle-antiparticle mixing}

\noindent
In the SM, there are no Flavor Changing Neutral Currents ($FCNC$) that directly convert $s\rt d$
or $b\rt s$, etc.
Thus, the SM descripton for mixing necessarily involves second-order weak-interaction
box diagrams as shown in Figs.~\ref{fig:k-mix} and \ref{fig:b-mix} above.  The process is mediated by
heavy virtual particles: in the case of $B^0$-$\bar{B}^0$ mixing, by virtual top-quarks and $W$-bosons.
If SUSY exists, virtual SUSY partner particles could also occur as virtual legs in the mixing box 
diagrams and, thus, cause differences between the  measured mixing frequencies and their SM prediction.

\begin{figure}[htb]
\centerline{  \includegraphics[height=0.28\textwidth,width=0.6\textwidth]{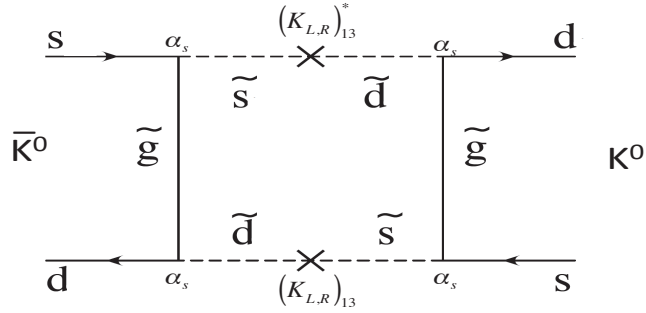}}
\caption{\footnotesize A potential SUSY box diagram contribution to $K^0$-$\bar{K}^0$
mixing. Here the vertical lines represent gluino propagators and the dashed lines squark
propagators. 
}
\label{fig:susy-mix}
\end{figure}  

A typical SUSY contribution to mixing is shown in Fig.~\ref{fig:susy-mix}, with virtual squarks and 
gluinos replacing the $W$s and up-type quarks in the SM diagrams shown in Figs.~\ref{fig:k-mix} and
\ref{fig:b-mix}.  
Here the $(K_{L,R})^*_{13}$ terms denote off-diagonal elements of a mixing matrix that accounts for the
possibility that squark flavors are not necessarily aligned with ordinary weak interaction flavors.
Here, unlike the SM diagrams, the vertices at the corners of the box are not due to weak interactions, but are
strong interactions. Thus, instead of being proportional the weak coupling $G_F$, they are proportional
to the QCD coupling $\alpha_s$,
which is much larger.  Thus, potential SUSY contributions to ordinary particle-antiparticle mixing are,
{\it a prior}, large.

Grossman provided a convenient formula that characterizes the relative strength of the SUSY and
SM contributions to $\Delta m_K$~\cite{grossman}:
\begin{equation}
\frac{\Delta m^{\rm SUSY}_K}{\Delta m^{\rm SM}_K}\sim 10^{4}\left(\frac{100~{\rm GeV}}{m_{\tilde{Q}}}\right)^2
\left(\frac{\Delta m^2_{\tilde{Q}}}{m^2_{\tilde{Q}}}\right)^2 {\mathcal Re}[(K_L)_{13}(K_R)_{13}].
\label{eq:dmsusy}
\end{equation}  
Here $m_{\tilde{Q}}$ is the squark mass and $\Delta m_{\tilde{Q}}$ is some measure of squark mass differences.

\begin{table}[htb]
\begin{center}\caption{\label{tbl:dm} Measured particle-antiparticle mixing frequencies
and their SM expected values. 
}
\begin{tabular}{l|c|c}\hline
Channel              &  $\Delta m$ (expt)      &  $\Delta m$ (theory) \\
                     &   (ps$^{-1}$)            &     (ps$^{-1}$)       \\
\hline\hline
$K^0$-$\bar{K}^0$    &  $0.00530 \pm 0.00001$   &   $0.00502 \pm 0.00051 $   \\
$B^0$-$\bar{B}^0$    &  $0.510 \pm 0.004$       &   $0.55^{+0.07}_{-0.05}$  \\
$B_s^0$-$\bar{B}^0_s$&  $17.69\pm 0.08$         &   $17.3\pm 2.6$      \\
\hline\hline
\end{tabular}
\end{center}
\end{table}

The measured values of the mixing frequencies for $K^0$, $B^0$ and $B_s^0$, taken from the
HFAG~\cite{HFAG} and PDG~\cite{PDG} averages, are compared to theoretical expectations~\cite{CKMfitter,yu}
in Table~\ref{tbl:dm}.  In all cases there is good agreement; the comparison is limited by the
precision of the theoretical calculations, which have errors in all three cases that are about $\pm$10\%.
The does not leave a lot of room for SUSY contributions per Eq.~\ref{eq:dmsusy}.  If we, somewhat arbitarily,
assume that the 10\%-level of agreement between experiment and SM predictions given in Table~\ref{tbl:dm} 
constrains the ratio in Eq.~\ref{eq:dmsusy} to be less than 0.3, we find that for a squark mass that would
have been accessible at the recent LHC run, {\it i.e.} $m_{\tilde{Q}}<1$~TeV, 
\begin{equation}
 \left(\frac{\Delta m^2_{\tilde{Q}}}{m^2_{\tilde{Q}}}\right)^2 {\mathcal Re}[(K_L)_{13}(K_R)_{13}]<\sim 0.003.
\label{eq:susy-constraint}
\end{equation}
This says that the venerable, 50 year-old,  measurements of $\Delta m_K$ place very severe constraints on
SUSY.  Either the squark masses must be highly degenerate, and very unlike  the quark masses which range
over nearly five orders of magnitude, or the SUSY squark flavors must align with the weak interaction
quark flavors to an extraordinary degree of precision, or some combination of the two.  Since there is
nothing in the SUSY theory that would naturally enforce such restrictions, this is called ``the Flavor
Problem,'' which is well known inside the SUSY community, but less well know generally.

To address this problem in a ``natural'' way, Nir and Raz devised a symmetry principle that restricted the
off-diagonal $(K_{L,R})$ matricx elements to higher order values of $\sin\theta_C$, thereby limiting
down-type squark contributions to particle-antiparticle mixing to acceptable values.  However, this
so-called ``horizontal symmetry'' necessarily requires up-type squark matrix elements to be of order
$\sin\theta_C$, which would make large SUSY contributions to the $D^0$-$\bar{D}^0$ mixing frequency
($\Delta m_c$); they predicted $\Delta m_c \simeq 0.1$~ps$^{-1}$~\cite{nir}.  Subsequent to the 
Nir-Raz paper, anomalously large  $D^0$-$\bar{D}^0$ mixing was observed by Belle and BaBar~\cite{dmix},
however this appears to be mostly in the real part of the mixing amplitude that relates the life-time difference
between the neutral $D$-meson eigenstates.
The HFAG group's average of the mass-difference measurements are within $\sim 2\sigma$ of zero~\cite{HFAG},
the 95\% confidence level upper limit is $\Delta m_c < 0.01$~ps$^{-1}$, an order of magnitude below
the Nir-Raz prediction.

\subsection{Time-dependent $CP$ violation asymmeties}

\noindent
The experimental precision of the mixing measurements discussed in the previous section far exceeds
that of the predictions based on SM theory.  Until the theoretical  precision is improved, there is
no pressing need for more precise experimental measurements.  However, the situation is reversed 
for the case of measurements of  $CP$ violating phases in mixing-induced interference effects.
For these, since QCD is $CP$ conserving, SM predictions for the $CP$ violating phases are not effected
by long-range effects or additional gluons etc., and are consequently less ambiguous and more precise. 

\begin{figure}[htb]
\centerline{  \includegraphics[height=0.45\textwidth,width=0.6\textwidth]{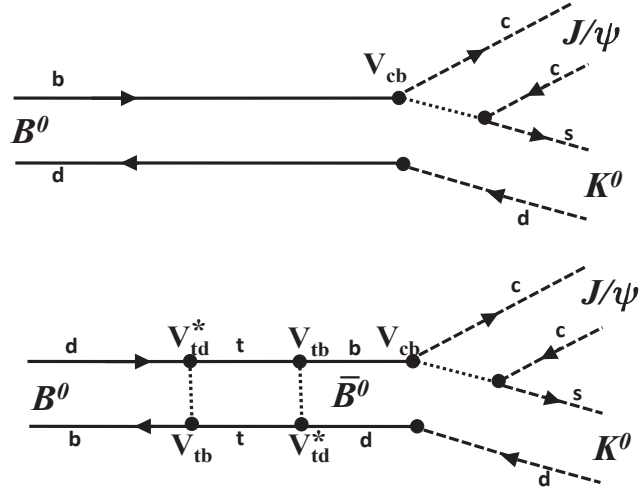}}
\caption{\footnotesize Quark flow diagrams that produce mixing-induced $CP$ violations.
The direct decay $B^0\rt K^0\jp$ {\it (upper)} interferes with $B^0\rt\bar{B}^0\rt K^0\jp$ decays {\it (lower)}.
Here the only flavor changing vertices with a non-zero $CPV$ phase is the $t\rt d$ transition.
The interference amplitude is proportional to $\sin 2\phi_1$, where $\phi_1$ is the phase of $V_{td}$.  
}
\label{fig:cp-mix}
\end{figure}

Mixing-induced time-dependent $CP$ violating asymmetries are due to the interference between the 
direct decay and of a $B^0$ meson to a $CP$ eigenstate such as $B^0\rt K^0\jp$ and the process where
tthe $B^0$ first transforms into a $\bar{B}^0$ and then decays to the same final state:
$B^0 \rt\bar{B}^0 \rt K^0\jp $, as illustrated in Fig.~\ref{fig:cp-mix}.  Here the 
$t\rt d$ vertices, which have a strength proportional to $V^{*}_{td}$, have
a $CP$ violating complex phase $\phi_1$.  In the SM, all the other vertices in the diagrams
shown in Fig.~\ref{fig:cp-mix} are real.
The interference between the two diagrams is proportional to $V^*_{td}V^{*}_{td}$ and has a
CPV phase of $2\phi_1$.

\begin{figure}[htb]
\centerline{  \includegraphics[height=0.4\textwidth,width=0.7\textwidth]{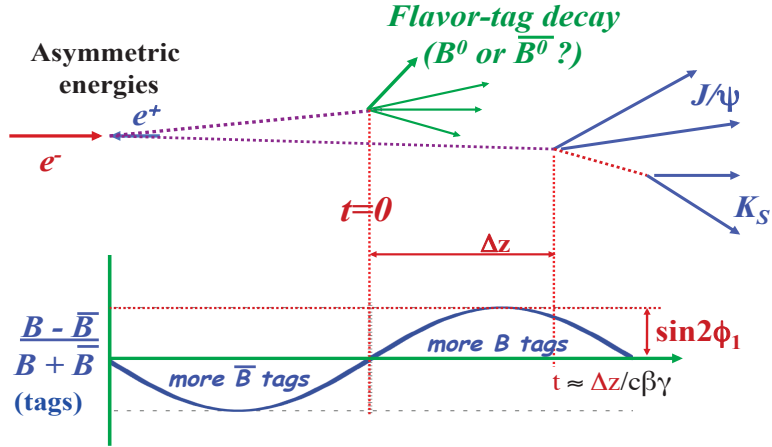}}
\caption{\footnotesize An illustration of the technique used to measure mixing-induced time-dependent $CPV$
asymmetries at the $B$-factories.
}
\label{fig:cp-meth}
\end{figure}

The measurement technique used at the $B$-factories is illustrated in Fig.~\ref{fig:cp-meth}.  A coherent $B^0\bar{B}^0$
meson pair is produced in an asymmetric $\ee$ collision.  After some time, one of the mesons ($B_{\rm tag})$ decays into
a flavor-specific final state, {\it i.e.,} a final state that reflects the $B$-flavor of the decaying meson. (These
types of decays occur more than 90\% of the time.) If the flavor of  $B_{\rm tag}$ can be tagged, {\it e.g.} by the
charge of a lepton from semileptonic decay or a charged kaon from the $b\rt c \rt s$ decay sequence, that ensures that the
accompanying meson, $B_{CP}$, has the opposite $B$-flavor at that time, which is set as $t=0$.  Then, as $B_{CP}$ evolves
with time, it starts to mix into its antiparticle state, the unmixed and mixed components of $B_{CP}$ interfere. The
interference can be seen when $B_{CP}$ decays into a $CP$ eigenstate.  The asymmetry between the number of times the
$B_{\rm tag}$ decay was a $B^0$ or a $\bar{B}^0$ as a function of the time that $B_{CP}$ decays, follows a
$\sin\Delta m_d t$ curve (both forward and backward in time) with amplitude $\sin 2\phi_1$, as sketched in the figure.
The relative time between the two decays is inferred from the measured $z$ position of each decay vertex. 
Figure~\ref{fig:cp-meth} shows the case for the $CP=-1$, $B_{CP}\rt K_S\jp$ decay mode.  Decays to $CP=+1$ final states,
such as $B_{CP}\rt K_L\jp$, have the opposite asymmetry. In the actual experiment, the amplitude of the asymmetry curve
is reduced from its ideal value of $\sin 2\phi_1$ by experimental dilution factors, primarily due to incorrectly tagged
$B_{\rm tag}$ decays.  The effects of these mistags are measured with high statistics $B^0$ semileptonic decay events in the
same data samples and are experimentally well understood.

\begin{figure}[htb]
\centerline{  \includegraphics[height=0.3\textwidth,width=0.75\textwidth]{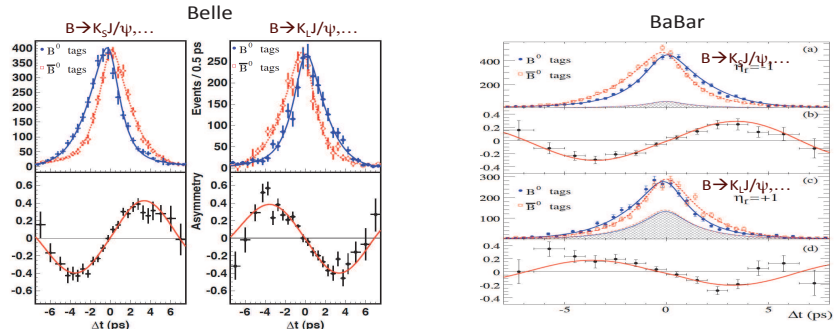}}
\caption{\footnotesize Belle {\it (left)} and BaBar {\it (right)} measurements of $CP$
violations for $B^0\rt K^0 (c\bar{c})$ decays. The upper plots show the time distributions for
numbers of $B^0$- and $\bar{B}^0$-tags and the lower plots show their time-dependent asymmetries.
}
\label{fig:cpv-full}
\end{figure}

This technique has been carefully developed and optimized by both the Belle and BaBar teams and it is now quite well
understood.  Results based on the full data samples from the two groups~\cite{belle_babar_phi1} are shown in 
Fig.~\ref{fig:cpv-full}. In these figures results from the $CP=-1$ and $CP=+1$  final states are shown
separately and they display opposite-sign asymmetries, as expected.  The measurements of $\phi_1$ are quite precise;
the current average value of the two groups' measurements is $\sin 2\phi_1 = 0.679\pm 0.020$ (corresponding to
$\phi_1=(21.38\pm0.79)^{\circ}$).  In both cases, the precision is still limited by statistics.  It is expected
that the precision could be improved by about a factor of two before measurements using this technique are 
systematics limited.  This is about the level of validity of the theoretical equivalence of the measured
asymmetry and $\sin 2\phi_1$. 

\subsection{BSM searches using mixing-induced $CP$-violating asymmetry measurements.}

\begin{figure}[htb]
\centerline{  \includegraphics[height=0.3\textwidth,width=1.0\textwidth]{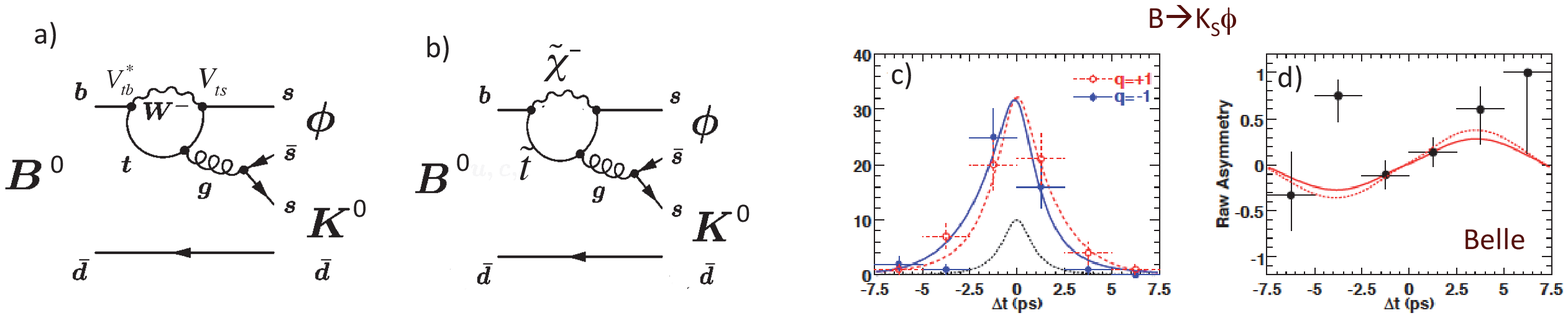}}
\caption{\footnotesize {\it a)} The SM penguin diagram for $B^0\rt K^0 \phi$ decays and
{\it b)} possible SUSY contributions.
Belle measurements of the time-dependent $B^0$ and $\bar{B}^0$-tag distributions and the $CPV$
asymmetry for $B^0\rt K_S \jp$ decays are shown in {\it c)} and {\it d)}, respectively.
}
\label{fig:ksphi}
\end{figure}

\noindent
Sensitive searches for signs of new, BSM physics can be made by applying the
above-described mixing-induced $CPV$ asymmetry measurement technique to rare
$B^0\rt CP$-eigenstate decays that are mediated by penguin diagrams.  An example
is the $B^0\rt K_S\phi$ decay mode, for which the SM leading order decay amplitude is the
penguin-mediated process shown in Fig.~\ref{fig:ksphi}(a).  The
CKM matrix elements involved here are $V^*_{tb}$ and $V_{ts}$, neither of which have 
a $CPV$ phase.  Thus, as in the case of $B^0\rt K^0\jp$ etc.,  the interference between
the direct $B^0\rt K_S\phi$ decay amplitude and the mixed $B^0\rt \bar{B}^0\rt K_S\phi$
amplitude all comes from mixing and will have the same $2\phi_1$ $CPV$ phase, and the
interference asymmetry will have the same $\sin 2\phi_1$ amplitude.

However, BSM theories that have new particles that can couple to $b$- and $s$-quarks can, in principle,
modify this process.  For example, in the case of SUSY, the $W$ and
$t$-quark in the SM process could be replaced by a squark and a chargino (the SUSY partner of the $W$),
as shown in Fig.~\ref{fig:ksphi}(b). In that case, the SUSY part of the decay amplitude would contribute a
different $CPV$ phase.  (SUSY has 44 non-trivial $CPV$ phases.)  Thus, a significant difference
between the mixing-induced $CP$-violating asymmetry in  $B^0\rt K_S\phi$ ($\sin 2\phi_1^{\rm eff}$) and that
in $B^0\rt K_S\jp$ would be a clear sign of new, BSM physics.  This is also the case for many
other penguin-mediated rare decays, such as $B^0\rt K_S\eta^{\prime}$, $B^0\rt K_S \pi^0$, etc.  

Such measurements were carried out by Belle and BaBar~\cite{belle-babar_penguin};
results from the Belle measurement
are shown in Figs.~\ref{fig:ksphi}(c) and (d).  The weighted average of the two groups'
measurements of $\sin 2\phi_{1}^{\rm eff}$ for $B^0\rt K_s\phi$ decays is 
$\sin 2\phi_1^{\rm eff}=0.39\pm 0.17$, which is $1.8\sigma$
below the SM expectation (of $0.679\pm 0.020$).  Results for other penguin processes have
similar precision.   The main current issue is statistics.  Penguin processes are rare: 
the branching fraction for $B^0\rt K_S\phi$ is a factor of a hundred smaller than
that for $B^0\rt K_S\jp$.   Thus, while the Belle measurements shown in Fig.~\ref{fig:cpv-full} 
include over 12K $B\rt K_S\jp$ events and more than 10K $B\rt K_L\jp$ events, the $K_s\phi$
measurements shown in Fig.~\ref{fig:ksphi} are based on only about 150 $B\rt K_S\phi$ events.
BelleII ultimately expects to accumulate a data sample 
that is $\sim$40 times larger than the Belle data sample.  With such a sample, $\sin 2\phi_1^{\rm eff}$
will be measured for $K_S\phi$ and many other penguin modes with a precisions of $\sim \pm 0.03$ 
for each mode.  These will severely test the SM.

The LHCb experiment is also challenging the SM with measurements of the mixing-induced phase $\phi_s$ in
$B_s\rt  \phi \jp$ decays. Since none of the CKM elements in the SM box diagram  for $B_s^0\rt\bar{B}_s^0$
have  a $CPV$ phase, the SM prediction for $\phi_s$ is that it should be very small.  This
was discussed at this meeting by U.~Uwer~\cite{uwer}.

\subsection{Generic new physics limits from mixing-induced $CP$-violation measurements}

\begin{figure}[htb]
\centerline{  \includegraphics[height=0.4\textwidth,width=0.85\textwidth]{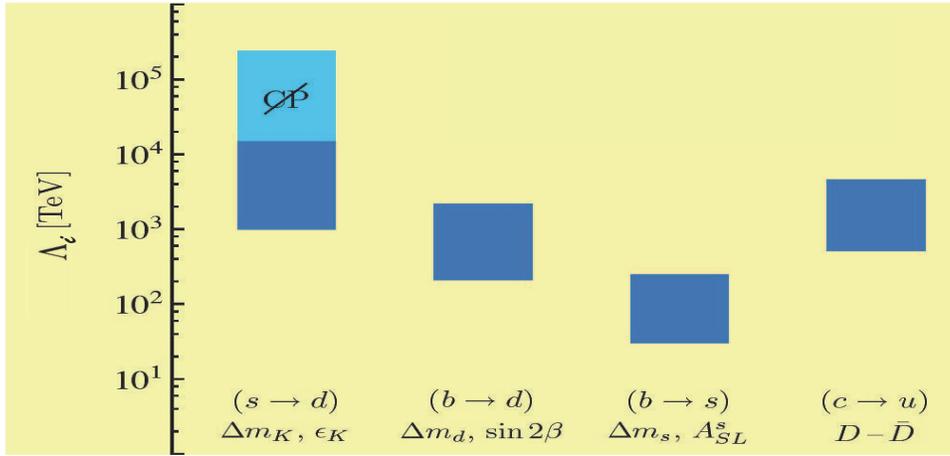}}
\caption{\footnotesize Range of limits on $\Lambda_i$ for different generic ({\it i.e.} $c_i=1$) new physics FCNC
processes channels for particle antiparticle oscillation and $CPV$ measurements.  Here $\epsilon_K$ is
the CPV parameter for the $K^0$, system, $\beta = \phi_1$, and $A^s_{\rm SL}$ is the limit on a $CPV$ asymmetry
in semileptonic $B_s^0$ decays.  (From M. Neubert's talk at EPS HEP 2011.) }
\label{fig:lambda}
\end{figure}

From dimensional analysis alone, the inclusion of new, BSM physics at a high mass scale $\Lambda_i$ 
that mediates the Flavor Changing Neutral Currents ($s\rt d$, $b\rt d$ $b\rt s$ and $c\rt u$) that are at play
in particle-antiparticle mixing could be described by an effective Lagrangian of the form
\begin{equation}
{\mathcal L}_{\rm eff} = {\mathcal L}_{\rm SM} + \frac{c_i}{\Lambda_i^2}O_i(FCNC),
\end{equation}
where $c_i$ is the coupling strength that depends on the details of the new physics theory,
and $O_i(FCNC)$ is the operator that produces the transition.  Thus, the influence of new physics
effects depends on $c_i/\Lambda_1^2$.  Figure~\ref{fig:lambda} illustrates the limits that 
current particle-antiparticle mixing measurements place on $\Lambda_i$ for the cases
where $c_i = 1$~\cite{neubert}.  For the $i= s\rt d$ transition, the allowed mass scales are
all above $10^3$ TeV.  For the other channels, the limits start nearer $10^2$~TeV, but still are
quite impressive.  Thus, as we noticed in the discussion associated with Eq.~\ref{eq:susy-constraint}
above, for new physics with mass scales that are accessible at current or future runs of the LHC, the
theories are tightly constrained to have very, very small values for the $c_i$ coefficients.   

\section{Search for signs of a charged Higgs}

\noindent
A feature common to many BSM theories is that they have Higgs sectors that are more complicated
than the single neutral Higgs scalar of the SM.  For example, in SUSY models there are
five Higgs scalars, including a doublet of charged Higgs particles $H^{\pm}$, and these all
have SUSY-partner higgsino fermions.  The search for charged Higgs particles is a major
activity of the high-$p_T$ LHC experiments.  

Models with charged Higgs doublets are classified into three types: in Type~I models,
the Higgs couples to up-type and down-type quarks with equal strength; in Type~II
models, the couplings to up-type and down-type quarks differ by a ratio that is commonly
expressed as $\tan\beta$;  Type~III models are all other cases.  Since many versions of
SUSY are Type~II models, these have been the most extensively studied.  

\begin{figure}[htb]
\centerline{  \includegraphics[height=0.4\textwidth,width=0.85\textwidth]{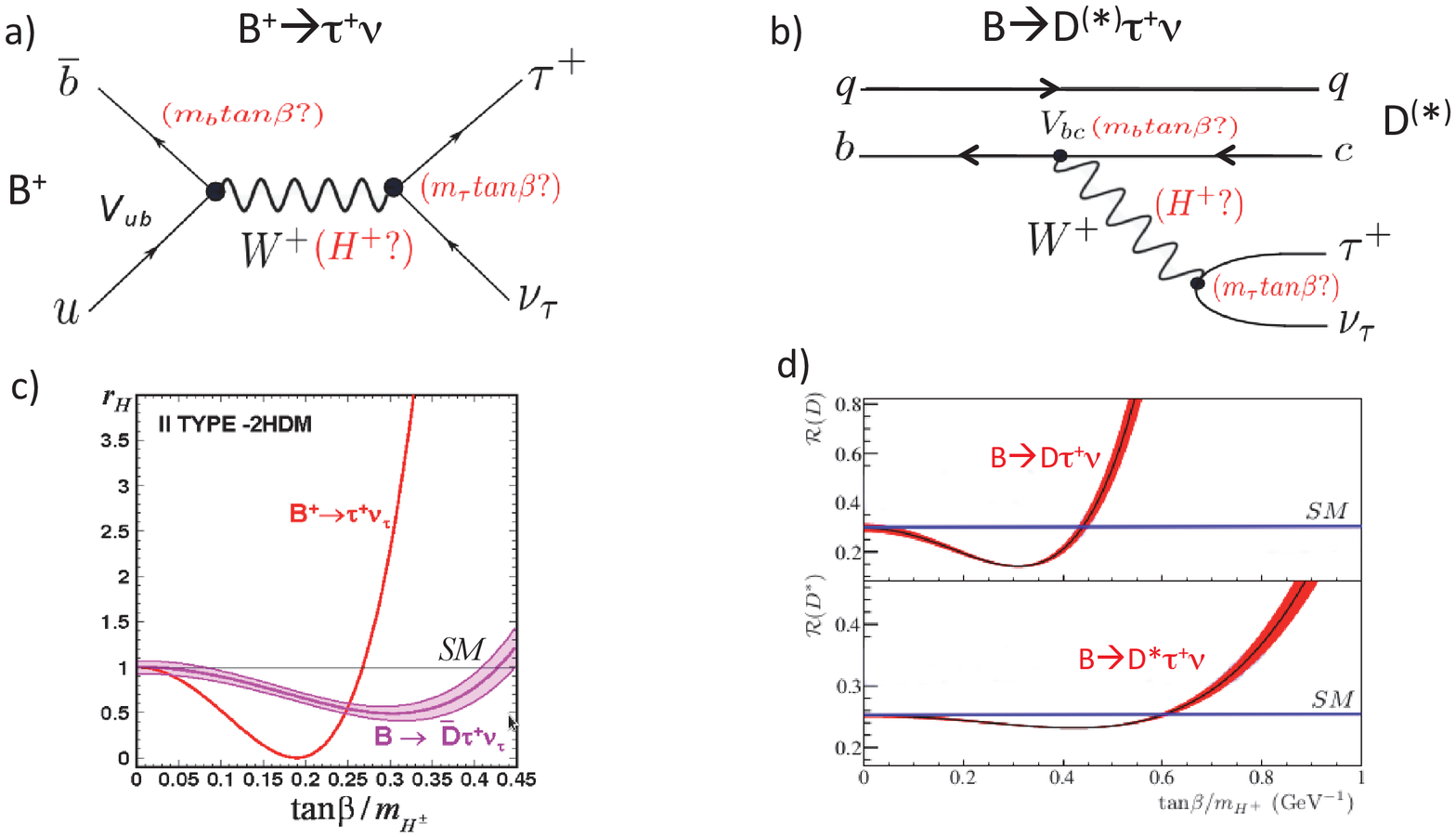}}
\caption{\footnotesize 
Quark line diagrams for {\it a)} $B^+\rt \tau^+\nu_{\tau}$ and {\it b)} $B\rt D^{(*)}\tau^+\nu_{\tau}$.
In the SM, the decays are mediated by virtual $W^{+}$ bosons; in BSM theories, virtual charged Higgs
bosons ($H^{\pm}$) can also contribute.  {\it c)} The ratio of the expected $B^+\rt\tau^+\nu_{\tau}$
branching fraction and its SM value ($r_H$) as a function of $\tan\beta /M_H$. {\it d)}
Corresponding ratios for $B\rt D\tau^+\nu_{\tau}$ ({\it upper)} and $B\rt D^*\tau^+\nu_{\tau}$
{\it (lower)}.   
}
\label{fig:chgd-higgs}
\end{figure}

Since Higgs couplings are proportional to mass, $B$ meson decays to final states containing 
$\tau$-leptons are most sensitive to possible effects from charged Higgs.
Figure~\ref{fig:chgd-higgs}(a) shows the quark-line diagram for purely leptonic $B^+\rt\tau^+\nu_{\tau}$
decay.\footnote{In this discussion, the inclusion of 
charge conjugate modes is implied.}

In the SM, this decay is mediated by a virtual $W^+$ boson that couples to the
$\bar{b}u$ vertex with a strength $V_{ub}G_F$; the SM expression for the branching fraction is
\begin{equation}
{\mathcal B}(B^+\rt \tau^+\nu_{\tau})=\frac{G_F^2 m_B m_{\tau}^2}{8\pi}(1-\frac{m_{\tau}^2}{m_B ^2})
f_B ^2|V_{ub}|^2\tau_B = (0.73^{+0.12}_{-0.07})\times 10^{-4},
\label{eq:taunu-SM}
\end{equation}
where $f_B$ is the $B^+$ decay constant calulated by Lattice QCD and $\tau_B$ is the $B^+$
meson lifetime.

If there is a Type~II charged Higgs with mass $m_H$,
it can also mediate this decay and would couple to the $\bar{b}u$ vertex with a strength
$m_b \tan\beta$ and to the $\tau\nu$ decay vertex with a strength $m_{\tau} \tan\beta$.
This modifies the branching fraction by a factor~\cite{hou} 
\begin{equation}
r_H = 1-\frac{m_B^2}{m_H^2}\tan^2\beta,
\end{equation}
which is shown as a function of $\tan\beta /m_H$ as a red curve in Fig.~\ref{fig:chgd-higgs}(c).

The same charged Higgs would also modify SM expectations for semileptonic
$B\rt D^{(*)} \tau\nu$ decays as shown in Fig~\ref{fig:chgd-higgs}(b). Here
the SM ``predictions'' are the measured branching fractions for $B\rt D^{(*)}\ell^+\nu$
($\ell^+ = \mu^+~{\rm or}~e^+$) scaled by  factor that reflects the reduced phase space
for $\ell^+ = \tau^+$.  The Type~II model charged-Higgs-induced modifications
for the $D\tau\nu$ and $D^{*}\tau\nu $ final states are different, and given by the ratios
to $D\ell^+\nu$ and $D^*\ell^+\nu$, $R_D$ and $R_{D^{*}}$, that are shown as a function of
$\tan\beta /m_H$ in the upper and lower sections of Fig.~\ref{fig:chgd-higgs}(d), respectively.   

\subsection{Experimental issues with $B\rt \tau\nu$ and $B\rt D^{(*)}\tau \nu$ measurements}

\begin{figure}[htb]
\centerline{  \includegraphics[height=0.4\textwidth,width=0.85\textwidth]{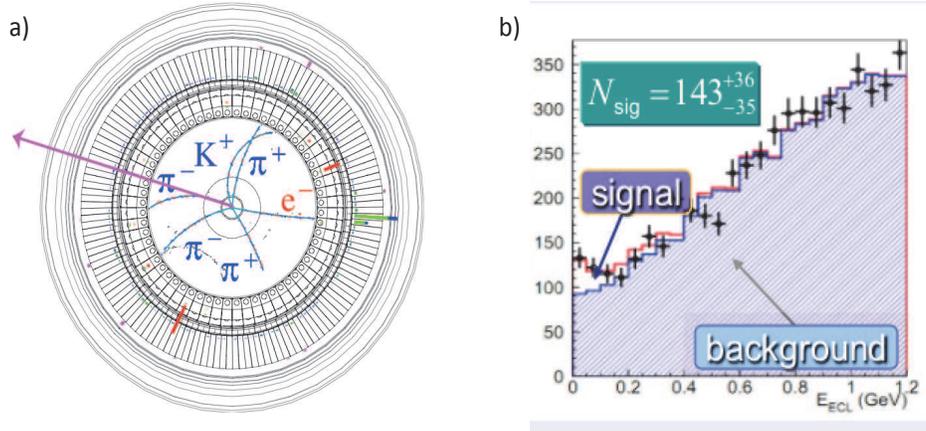}}
\caption{\footnotesize {\it a)} An event display of a $B^-\rt\tau^-\bar{\nu}$ event candidate in Belle. 
Here a $B^+$ is fully reconstructed as discussed in the text and the $\tau$ decays via the 
$\tau^-\rt e^-\nu\bar{\nu}$ mode.  Here the purple arrow indicates the direction of the missing
momentum in the event.  {\it b)} The distribution of unassigned energy in the CsI
calorimeter for $B\rt \tau\nu$ candidate events.  The small excess over background below 200~MeV is 
the signal for $B\rt\tau\nu$.
}
\label{fig:evt-disp}
\end{figure}

\noindent
Experimentally, measurements of $B$ meson decay channels that contain a $\tau$ and a $\nu_{\tau}$
are challenging.  Final states have at least two neutrinos and, thus, missing mass techniques
commonly used to study single neutrino final states associated with $e^-\nu$ and $\mu^-\nu$ lepton pairs
are not applicable. The BaBar and Belle experiments use the ``tagged $B$'' technique that exploits the
fact that the $B$ mesons produced in $\ee$ collisions at the $\Upsilon (4S)$ are produced in $B\bar{B}$
pairs, with no accompanying particles.  Thus, if one $B$ meson decay is completely reconstructed,
one knows with confidence that any remaining particles in an event must be decay products from
 the accompanying $\bar{B}$. For the $B^-\rt\tau^- \bar{\nu}$ measurement, Belle selects events that
contain one fully reconstructed $B^+$ meson and a single accompanying charged track, which is
potentially a lepton from $\tau^-\rt \ell^-\nu\bar{\nu}$ or a $\pim$ from $\tau^-\rt \pim \bar{\nu}$~\cite{belle_tau}.  
An event display of a candidate $B^-\rt\tau^-\nu$ decay in Belle is shown in Fig.~\ref{fig:evt-disp}(a),
where a $B^+\rt\bar{D}^0 \pip$ decay, with $\bar{D}^0\rt K^+\pim\pipi$, is reconstructed in the
tracking system, along with a single, well identified electron.  
Many other $B^-$ decay channels produce a single track in the detector.  However, these also
contain other neutral particles, such as $\gamma$-rays, $\piz$s, $K_L$ mesons, etc. Usually,
these particles deposit sgnificant amounts of energy in the CsI calorimeter that surrounds the
Belle tracking system, covering $\sim$85\% of the total $4\pi$ solid angle.  The signature for
$B^-\rt \tau^-\bar{\nu}$ is an excess of events with small excess energy deposit in the calorimeter,
as shown in Fig.~\ref{fig:evt-disp}(b).  Here the background level is established by studying events
in which a tagged $B$ is accompanied by a meson that decays via the semileptonic $B^-\rt D^{*0}\ell^- \bar{\nu}$
decays, and other well understood control samples.  The $B\rt D^{(*)}\tau\nu$ event selection, described
in Ref.~\cite{belle_dtau}, is similar. Since the probability for fully reconstructing the accompanying
$B$ meson is very low, $\sim 0.25\%$, this method has very low efficiency, but it is best that one can do.
These measurements can only be done at an $\ee$ $B$-factory operating at the $\Upsilon(4S)$.    

The 2013 PDG world average $B\rt\tau\nu$ branching fraction
is $(1.05\pm 0.25)\times 10^{-4}$~\cite{PDG}, which is higher than the SM value, given above in
Eq.~\ref{eq:taunu-SM}, by $\sim 1\sigma$.  Results from Belle~\cite{belle_dtau} and
BaBar~\cite{babar_dtau} on $B\rt D\tau\nu$ and $B\rt D^*\tau\nu$ range from $1\sigma$ to
$\sim 2.5\sigma$ above the SM expectations.  These discrepancies are intriguing, but not
significant enough to make any claims.  

\begin{figure}[htb]
\centerline{  \includegraphics[height=0.4\textwidth,width=0.85\textwidth]{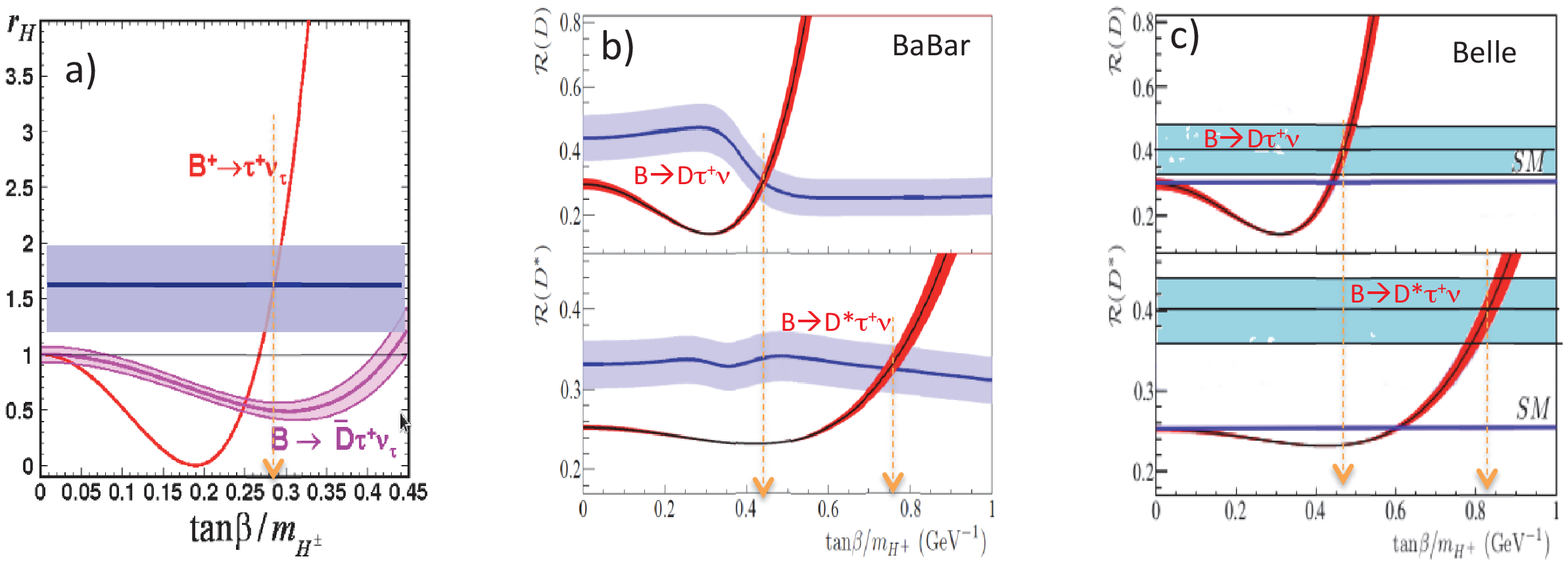}}
\caption{\footnotesize 
{\it a)} The blue band shows combined BaBar and Belle results on  $B^-\rt\tau^-\bar{\nu}$ compared with
expectations from the SM (horizontal line at $r_H=1$) and the Type~II Higgs doublet model. The
measured value is about $1\sigma$ above SM expectations. {\it b)} BaBar results on $B\rt D\tau\nu$
{\it (upper)} and $D^+\tau\nu$ {\it (lower)}.  Here for both modes the measured values are about
$2\sigma$ above SM expectations. {\it c)} Corresponding results from Belle.
}
\label{fig:taunu-results}
\end{figure}  

The situation is summarized in Fig.~\ref{fig:taunu-results}, where current experimental results,
shown as blue bands, are superimposed on the curves indicating expectations from a Type~II
Higgs doublet model shown above in Fig.~\ref{fig:chgd-higgs}.  The locations of the crossing
points of the experimental bands with the red expectation curves should locate the value
of $\tan\beta /m_H$.  As can be seen in the figure, the preferred $\tan\beta /m_H$ values for
each mode are inconsistent: $\sim 0.28$ for $\tau\nu$, $\sim 0.45$ for $D\tau\nu$, and
$\sim 0.8$ for $D^*\tau\nu$.  These discrepancies are undoubtably are due to the large statistical
errors -- there is no compelling evidence in the data for a charged Higgs to begin with.  However, one could
imagine a situation  where, with 40 times more data as expected for BelleII, the results settle
on the current central values but with error bars that are four or five times smaller and, at the same time,
LHC experiments report signals for a charged Higgs.  In that case, not only would the $B$ decay
measurements provide additional compelling evidence for some BSM process, but also clear diagnostic
evidence  that rules out its interpretation as a Type~II charged Higgs.

\section{Comments}

\noindent
In this talk I intentionally avoided giving a shopping list of physics topics\footnote{More
comprehensive coverage of the broad range of topics that are on the agenda for near-future,
intensity-frontier flavor physics experiments are provided in Refs.~\cite{belleii_physics}
and~\cite{superb_physics}, and references cited therein.} 
and, instead,
tried to provide a few detailed and concrete examples of how flavor physics measurements
place tight constraints on any proposed model for physics beyound the Standard Model, and
also how future precision measurements of rare processes can probe for new physics at mass
scales that are far above those that will ever be accessed by the LHC.  I described a scenario
that illustrates how flavor physics measurements will provide essential diagnostic information
that can help classify the ``DNA'' of a charged Higgs candidate that might be observed by LHC
experiments.  The flavor physics measurements are an essential component of new, BSM physics 
physics searches.  

\section{Acknowledgements}

\noindent
I thank the organizers for inviting me to give this talk and congratulate them
on another interesting and successful Bormio meeting.  This
work was supported in part by the Korean National Research Foundation (NRF)
Grant No.~20110029457.

\end{document}